# A hit-and-run Giant Impact scenario

Andreas Reufer[1], Matthias M. M. Meier[2,3], Willy Benz[1] and Rainer Wieler[2]

[1] Physikalisches Institut & Center for Space and Habitability, Universität Bern, Sidlerstrasse 5, 3012 Bern, Switzerland

[2] Institut für Geochemie und Petrologie, Eidgenössische Technische Hochschule Zürich, Clausiusstrasse 25, 8092 Zürich, Switzerland

[3] Department of Geology, Lund University, Sölvegatan 12, 22362 Lund, Sweden

*Abstract*

*The formation of the Moon from the debris of a slow and grazing giant impact of a Mars-sized impactor on the proto-Earth (Cameron & Ward 1976, Canup & Asphaug 2001) is widely accepted today. We present an alternative scenario with a hit-and-run collision (Asphaug 2010) with a fractionally increased impact velocity and a steeper impact angle.*

## 1. **Introduction**

Hydrodynamical simulations have identified a slow, grazing impact in being able to reproduce the Moon's iron deficiency and the angular momentum of the Earth-Moon-system. But in this canonical scenario, the Moon forms predominantly from impactor material, thus contradicting the Moon's close geochemical similarity to Earth. Furthermore, due to the slow impact velocity, only limited heat input is provided for the aftermath of the collision. Post-impact mechanisms (Pahlevan & Stevenson 2007) required to match the impact scenario with the compositional observations, depend on the thermal conditions in the post-impact debris disk. We show that a new class of hit-and-run collisions with higher impact velocities and a steeper impact angles is also capable of forming a post-impact debris disk from which the Earth's Moon can later form, but leads to a much hotter post-impact debris disk. Furthermore, the ratio of target body material in the debris disk is considerably larger, compared to the canonical scenario. This new class of impacts was previously rejected due to the limited resolutions of early simulations (Benz 1989).

## 2. Methods

We use a 3D SPH code with self-gravity but without strength (Reufer 2011) similar to previous work modelling the Giant impact (Benz 1989, Canup & Asphaug 2001). Smoothed Particle Hydrodynamics (Monaghan 1992) represents matter in a Lagrangian way by representing mass as individual particles. This allows to track not only the type of material but also the origin of the material. For the equation of state, we use ANEOS (Thompson & Lauson 1972) for iron and water ice, and M-ANEOS (Melosh 2007) for the silicate layers composed of quartz.

Initial target and impactor are set up, by starting with isentropic vertical profiles for desired masses and composition. The physical quantities of those profiles like density and specific energy are then mapped onto spheres of SPH particles. These spheres are relaxed until the residual particle velocities fall below 1% of the body's escape velocity, the latter being on the order of magnitude of the impact velocity. Individual simulations are then setup by placing relaxed spheres onto the orbit required for a collision with the desired impact angle and velocity and setting the initial velocity vectors accordingly. The total number of particles in the simulations varies between 500k to 550k, so that a post-impact disk of a Moon mass is represented by roughly 10k particles. This is similar to recent high-resolution simulations of the giant impact and good enough to determine the composition and thermal state of the post-impact disk.

Simulations are post-processed the same way as in previous work (Benz et al. 1986, Canup & Asphaug 2001, Canup et al. 2001, Canup 2004, Canup et al. 2012): First the particles of the central body are found with the following algorithm: Particles near the center of mass of all particles are taken as a first guess for the central body. New particles are added to the existing body, if they are gravitationally bound to it, but not in an orbit around it. The surface of the central body is then given by the particles, which have a density near the zero-pressure density of the material. For the remaining particles the orbits are then determined relative to the central body. If the periapsis lies above the central body's surface and the orbit is closed, a particle is considered a part of the debris disk. Otherwise particles are either re-impacting the central body or escaping the system.

While in the canonical case, the orbital time scale of the material in the debris disk is on the order of only a few hours, the increase in impact velocity in our simulations also leads to an increase in the orbital time scale of the disk material. Therefore we integrate the simulations to roughly 50h after the actual impact.

## 3. **Theory**

The collisional parameters for the giant impact considered so far have, however, been limited to low-velocity collisions at or only slightly above mutual escape velocity between target and impactor. The Earth-Moon system is not believed to have lost more than 10% of its initial angular momentum between the giant impact and today (Canup et al. 2001). In low-velocity collisions, very little mass and therefore also very little angular momentum is lost, compared to the total mass and angular momentum before the collision. Hence if impact velocity and angular momentum of the collision are fixed values, for a given impactor and target composition only one degree of freedom remains in the form of the product of the sine of impact angle and the impactor mass. Previous work therefore focussed on finding the optimum mass ratio between impactor and target. Recent work suggests a mass ratio of 9:1 and a total mass of 1.05 $M_E$ (Canup 2004). Both the impactor and the target are assumed to be differentiated bodies with a 30wt% iron core and a 70wt% silicate mantle. In these low-velocity collisions, the impactor loses kinetic energy during its grazing collision with the target, before it is dispersed into a disk around the target. The resulting proto-lunar disk is therefore mainly composed of impactor material. We will call this the “canonical scenario”.

When the assumption that no mass is lost is dropped however, the collisional angular momentum is no longer tightly constrained, as lost mass also carries away angular momentum. The total collisional angular momentum can therefore be considerably higher than the final angular momentum in the Earth-Moon system. With this additional degree of freedom , new regions in the collision parameter space become feasible.

Apart from the disk mass, another interesting quantity is the origin of the material which ends up in the proto-lunar disk, especially for the silicate part.

We call the fraction of target silicate to total silicate material in the disk

$$f_T = (M^{silc}_{targ} / M^{silc}_{tot})_{disk} \quad (1)$$

where $M^{silc}_{targ}$ and $M^{silc}_{tot}$ denote the mass of the silicate fraction of the disk derived from the target, and the total disk mass, respectively. If we define a similar target-derived silicate fraction for the post-impact Earth, we can deduce a deviation factor

$$\delta f_T = ( M^{silc}_{targ} / M^{silc}_{tot} )_{disk} / ( M^{silc}_{targ} / M^{silc}_{tot} )_{post\text{-}impact\ Earth} - 1 \quad (2)$$

which directly reflects the compositional similarity between the silicate part of the proto-lunar disk and the silicate part of the post-impact Earth.

Isotopic measurements show (Wiechert et al., 2001 & Zheng et al., 2012) a strong isotopic similarity between the silicate fractions of today's Moon and Earth. Assuming isotopic heterogeneity of the pre-impact bodies, this suggests that either the material of the bodies mixed during the collision or re-equilibrated their isotopic signatures after the collision. Either scenario is represented by a $\delta f_T \sim 0$ between today's Earth and the Moon. The value of $\delta f_T$ right after the impact thus serves as a starting point, from which a re-equilibration mechanism leads to todays value of $\delta f_T \sim 0$.

In a typical simulation of the canonical scenario, only about 30% of the disk material and 90% of the material of the post-impact Earth is derived from the target (the proto-Earth) respectively (Canup 2004), yielding a $\delta f_T$ of -67%.

## 4. **Results**

The new class of collisions presented here falls into the broad regime of slow hit-and-run collisions (Asphaug et al. 2006) with impact velocities between 1.20-1.40 $v_{esc}$. Hit-and-run occurs up to half the time for collisions with impact velocities in this range. Because of the higher impact velocities in this type of collisions, substantial mass and angular momentum can be lost in the process. Therefore, the initial angular momentum is less constrained and can be considerably higher than the post-impact 1.0-1.1 $L_{E\text{-}M}$ angular momentum of the Earth-Moon-system. The higher impact velocities used in these simulations are also encouraged by more recent models of terrestrial planet formation (O'Brien et al. 2006). In hit-and-run collisions, a significant part of the impactor escapes, so that the disk fraction is enriched in target-derived materials compared to the canonical case. Figure 1a shows four consecutive snapshots of such a hit-and-run collision. While the overall characteristics of the collision resemble the canonical scenario, here a considerable part of the impactor is ejected.

In the new class of giant impact simulations presented in the following paragraphs, for the first time a significantly higher fraction of the material constituting the disk is derived from the Earth's mantle. Table 1 shows a selection of around 60 simulations performed by us. A canonical reference run (cA08) uses initial parameters and conditions similar to those used in the canonical scenario (Canup 2004), successfully reproducing an iron-depleted proto-lunar disk massive enough to form a Moon. For each of our runs, a final Moon mass is calculated using the disk mass and the specific angular momentum (Kokubo, Ida & Makino 2000). We employed three different impactor types with different compositions: chondrictic impactors with 70wt% silicates and 30wt% Fe ($v_{esc}$ = 9.2km/s, c-runs), iron-rich impactos with 30wt% silicates and 70wt% Fe ($v_{esc}$ = 9.3km/s, f-runs) and icy impactors with 50wt% water ice, 35wt% silicates and 15wt% Fe (vesc = 8.9km/s, i-runs). Note that the scaled impact velocity $v_{imp}/v_{esc}$ determines the type of collision, as similar-sized collisions in the gravity regime are self-similar (Asphaug 2010, Leinhardt & Stewart 2012). Initial temperatures of the iron cores are between 4000-5000K, for the silicate layers between 1600K-2200K and around 300K for water ice layers. As mentioned before the layers are isentropic and the temperature therefore varies with depth.

## 5. Discussion

The ratio of target- vs. impactor-derived material that ends up in the proto-lunar disk is mainly defined by the geometry of the collision during the very early phase when the impactor is accelerating target material. This can be seen in Figure 1b, where the particles which later end up in the disk are highlighted in bright colours. In the canonical scenario, the impactor grazes around the target's mantle and is deformed. Due to the low impact velocity, material supposed to end up in orbit around the Earth must not be decelerated too strongly in order to retain enough velocity to stay in orbit. This is only achieved for the parts of the impactor mantle most distant to the point of impact, and some minor part of the target's mantle. But if impact velocity is increased from 1.00 (cA08) to 1.30 $v_{esc}$ (cC01), parts from deeper within the target mantle receive the right amount of energy for orbit insertion, while the outer regions of the target mantle, retain too much velocity and leave the system, thereby removing mass and angular momentum. Both processes work towards increasing the target material fraction in the proto-lunar disk. While in run *cB04* only ~ 10% of the initial angular momentum is removed, ~ 45% are removed in run cC06.

We have found that collisions with an impact angle of 30 - 40° and impact velocities of 1.2 – 1.3

$v_{esc}$ are successful in putting significant amounts of target-derived material into orbit, when using differentiated impactors with a chondritic iron/silicates ratio (30wt% Fe, 70wt% silicates) and masses between 0.15-0.20 $M_E$. Some runs in this regime show an iron excess of > 5wt% in the proto-lunar disk and are rejected, as in previous work (Canup 2004). While none of the runs done so far provide a “perfect match” in terms of the constraints from the actual Earth-Moon-system, several simulations come close to that. The best runs coming close to matching the constraints (cC03 and cC06) are obtained using impact angles of 32.5° and 35° and velocities of 1.25 and 1.20 $v_{esc,}$, resulting in 54% and 56% of the silicate material deriving from the target, and $\delta f_T$ thus increasing to -35% and -37% compared to -66% in the reference run of the canonical case. While the satellite masses match well (1.01 and 1.24 $M_L$), the angular momentum of the runs is somewhat too high (1.28 $L_{E-M}$). This should, however, be contrasted with other runs, e.g. cB03, where the reduction of the impactor mass to 0.15 $M_E$ results in a similar disk-composition ($\delta f_T$ = -33%), but also a lower Moon mass (0.53 $M_L$) and a smaller angular momentum of 1.06 $L_{E-M}$.

As collision geometry predominantly determines the fraction of target material in the proto-lunar disk, altering the size of the impactor by density changes should also change the target material fraction in the disk. A denser impactor with the same mass delivers the same momentum, while reducing “spill-over” of impactor material into the disk, as it can be seen in figure 1b. To include such a high density, iron-rich impactor in this study is also motivated by the work of Asphaug 2010, where the population of second-largest bodies in a planet-forming disk becomes slowly enriched in iron through composition-changing hit-and-run collisions. We investigated impactors with 50%wt and 70wt% iron core fractions, respectively. With a 0.2$M_E$ impactor at 30° impact angle and 1.30 $v_{esc}$ impact velocity, the target material fraction $f_T$ increases from 57% ($\delta f_T$ = -34%) in the “chondritic” run (cC01), to 64% ($\delta f_T$ = -28%) in the 50wt% iron core run (fA01), and up to 75% ($\delta f_T$ = -19%) in the 70wt% iron core run (fB06). But at the same time, the iron content of the disk increases to values incompatible with lunar data, and also the bound angular momentum increases to unrealistic values. Apparently, reducing the "spill-over" also reduces the lost mass and therefore the lost angular momentum.

We also looked into less dense, but still fully differentiated impactors with a composition of 50wt% ice, 35wt% silicate and 15wt% iron, typical for small bodies accreted in regions of the solar system beyond the snow-line. In these runs, the efficiency of putting material into orbit is reduced compared to runs with denser impactors, though the fact that there is less impactor silicate material available to end up in the disk actually raises the target material fraction to 81%, and $\delta f_T$ to -10% in the case of an $0.2M_E$ impactor hitting at 30° impact angle and 1.30 $v_{esc}$ impact velocity (cC01). However, in most runs, the resulting silicate portion of the disk is not massive enough to later form the Moon, although also one run with a silicate portion of 0.73 $M_L$ (iA10) was found.

A remarkable difference between the new simulations presented here and the canonical model, relevant to the re-equilibration model, is the thermal state of the Earth after the collision. The higher impact velocity leads to stronger shocks and therefore to more intense impact heating. Figure 2 shows cuts through the post-impact Earth ~ 58h after the collision, with colour-coded temperature. Temperatures are shown as estimated by ANEOS for the specific energy as it being integrated during the course of a simulation. Note that the values should only be regarded as order of magnitude estimates as heating due to shocks is dependent on how well shocks are resolved, which is resolution dependent. In the canonical case (Figure 2, left panel), only a thin blanket of material raining back from the disk experiences strong heating, while most of the mantle is only moderately heated above the initial average temperature of 2000 K. In the slow "hit-and-run" scenario (Figure 2, right panel), the shock is stronger, and a substantial part of the mantle is heated to temperatures above 10'000 K. Beside possible implications for the early history of the Earth, this also constitutes a different starting point for the re-equilibration model (Pahlevan & Stevenson 2007). While in the canonical model, the thin hot blanket may inhibit convection within the Earth, and thereby isotopic re-equilibration at early stages, the much thicker hot layer which extends deep into the mantle in our model probably simplifies exchange of material between the deep interior of the post-impact Earth and the hot silicate vapour atmosphere. It is important to note, that the post-impact temperatures depend not only on the magnitude of shock heating experienced during the collision, but also on the initial temperature profiles of the pre-impact bodies and the numerical method employed. The absolute temperatures obtained in this study therefore should only be used to qualitatively compare the effect of different impact parameters on the impact heating. Also note that our model neglects strength and therefore leads to a slightly changed attenuation of the shock wave and therefore to a changed shock heating.

Our results not only relax the required efficiency of re-equilibration by starting with a disk with a higher target material fraction, but the resulting thermodynamical state might actually create a more favourable starting point for re-equilibration.

Another important difference to the canonical scenario is the dynamical state of the proto-lunar disk. In both scenarios, material is ejected on ballistic trajectories after impact and forms an arm-like structure. Angular momentum is transferred from the inner to the outer part, thus circularising the orbit of the outer part of the arm while the inner part re-impacts the Earth. In the "hit-and-run" scenario, this arm structure persists for a shorter time compared to the canonical case, transferring less angular momentum and leaving the material on more eccentric orbits. The dynamical evolution of such an eccentric disk and the consequences for Moon formation have yet to be studied. Also, the effects of pre-impact rotation of both the target and the impactor will have to be studied in future work.

## 5. **Conclusions**

In summary, our new giant impact scenario based on a "hit-and-run" collision at higher impact velocity (1.20 – 1.25 $v_{esc}$) and a steeper impact angle (30 – 35°) than assumed in the canonical scenario provides an alternative scenario for the giant impact, with a different starting point for lunar accretion in the post-impact debris disk. The thermal state of this disk is considerably hotter compared to the canonical scenario.

## 6. Acknowledgements

We thank Maria Schönbächler, Kaveh Pahlevan and Vera Fernandes for discussions and Jay Melosh for providing us with the M-ANEOS package. Andreas Reufer and Matthias M. M. Meier were both supported by the Swiss National Science Foundation. All calculations were performed on the ISIS2 cluster at University of Bern.

**Figure 1**

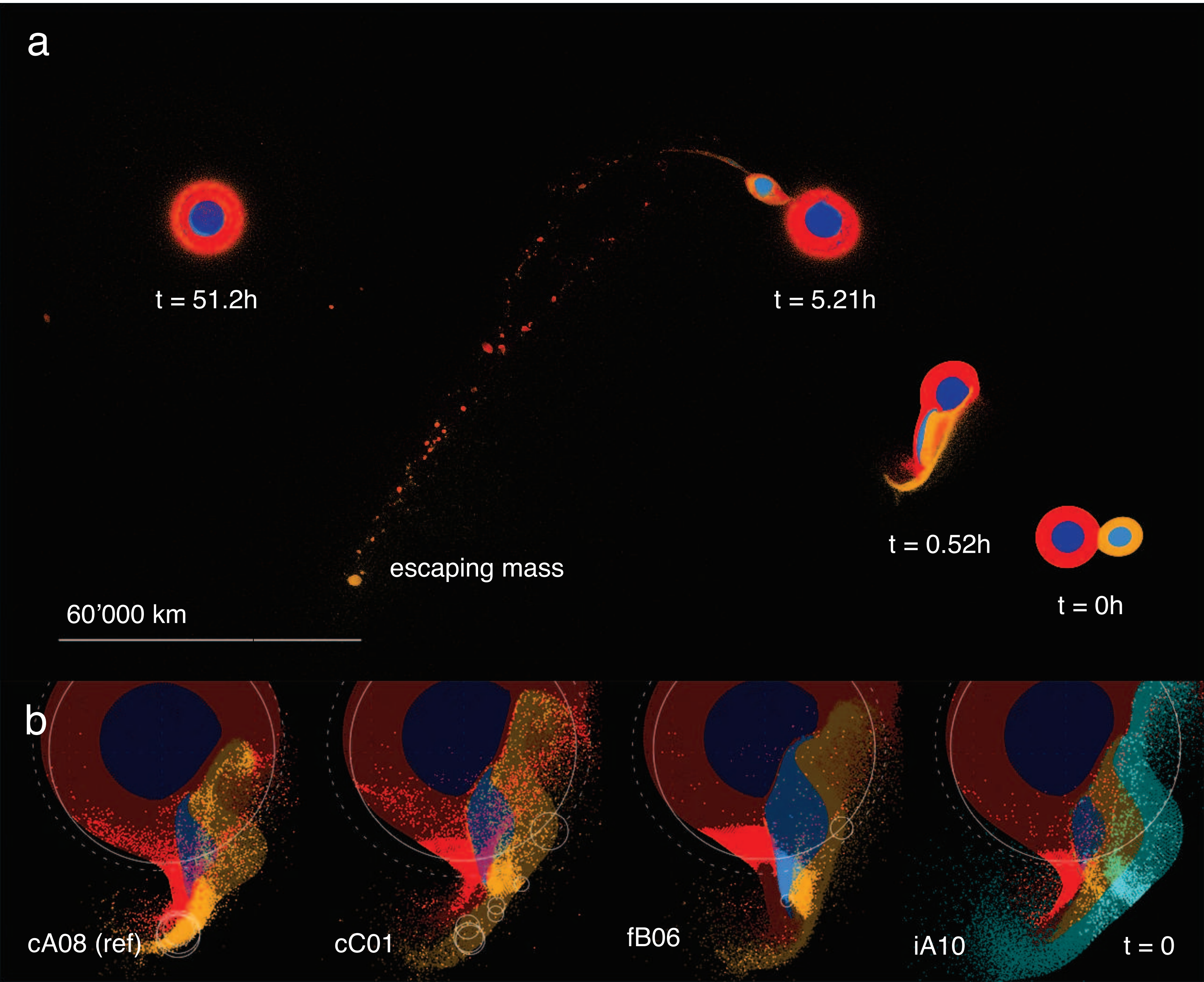

Figure 2

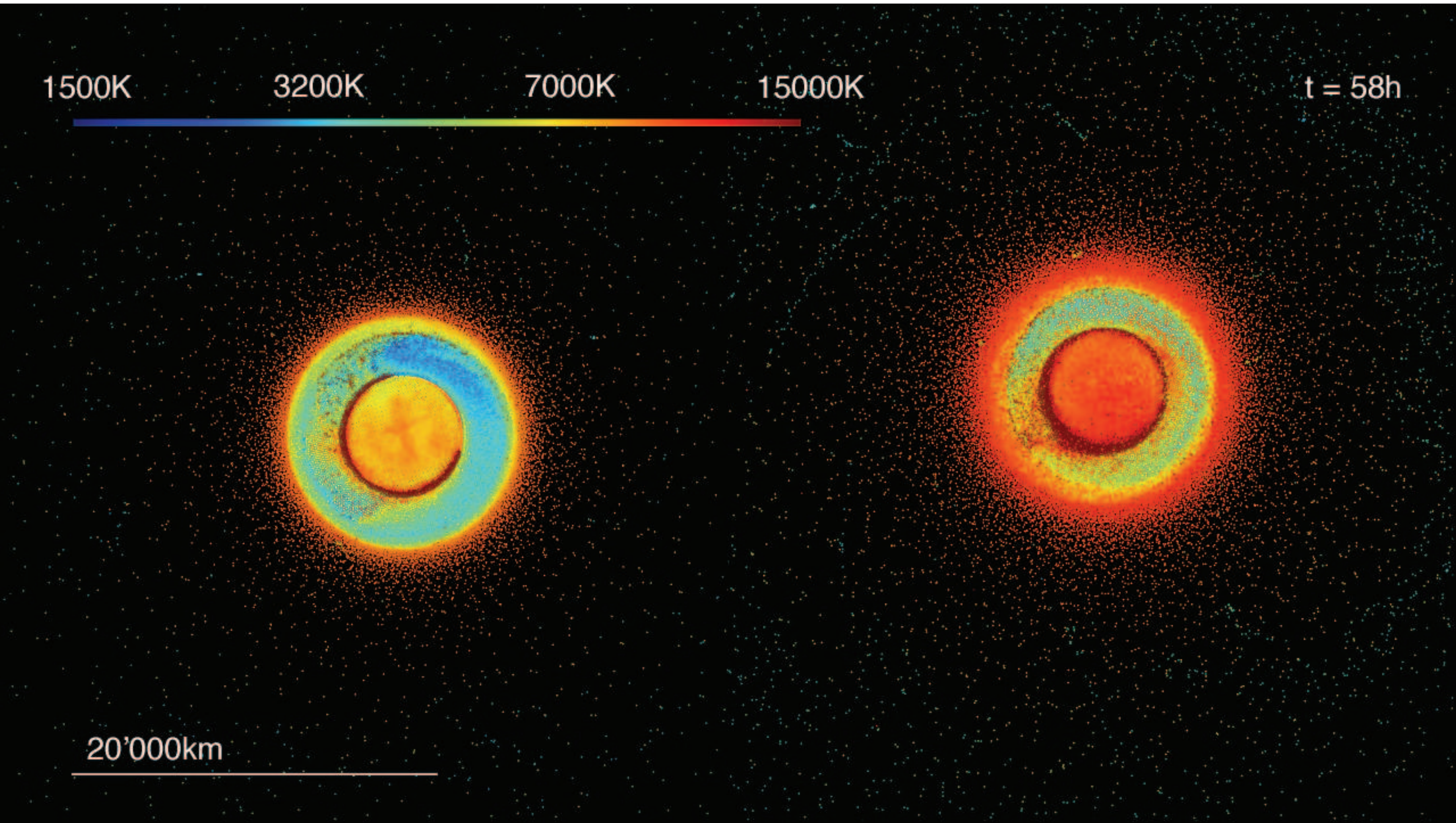

**Figure captions**

Figure 1a: Five snapshots from the 30° impact angle and 1.30$v_{esc}$ impact velocity case (cC06) showing cuts through the impact plane. Colour coded is the type and origin of the material. Dark and light blue indicate target and impactor iron; Red and orange show corresponding silicate material. The far right shows the situation at the time of impact. At 0.52h, it can be seen how the impactor ploughs deep through the targets mantle and pushes considerable amount of target material into orbit. A spiral arm of material forms and gravitationally collapses into fragments. The outer portions of the arm mainly consist of impactor silicates and escapes due to having retained a velocity well above escape velocity. The silicate fragments further inward are stronger decelerated and enter eccentric orbits around the target. The impactor's iron core also looses much of its angular momentum to the outer parts of the spiral arm and re-impacts the proto-Earth.

Figure 1b: The origin of the disk material highlighted, half a collisional timescale ( $(R_{imp} + R_{tar}) / v_{imp}$ ) after impact. In the grazing reference case (cA08), the majority of the proto-lunar disk originates from a spill-over of the impactor. In the head-on cases (cC01, fB06, iA10), much more material comes from the target mantle, being pushed out into orbit by the impactor core. Colours are identical to figure 1. Turquoise on the right shows water ice for the icy impactor case iA10.

Figure 2: Comparing post-impact temperatures of the proto-Earth between the grazing reference simulation left (cA08) and the head-on case on the right (cC06). Colour coded is temperature in K in logged scale. The initial average temperature before the impact inside the target mantle is ~ 2000K.

**Table 1**

| run | $M_{tar}$ [$M_E$] | $M_{imp}$ [$M_E$] | θ [°] | $v_{imp}$ [$v_{esc}$] | impactor [wt% Fe/$SiO_2$/$H_2O$] | $\delta f_T$ | $f_T$ | $M_{moon}$ [$M_L$] | $L_D$ [$L_{E\text{-}M}$] | $M_D$ [$M_L$] $SiO_2$ | Fe | $H_2O$ | $L_{bound}$ [$L_{E\text{-}M}$] | $L_{imp}$ [$L_{E\text{-}M}$] |
|---|---|---|---|---|---|---|---|---|---|---|---|---|---|---|
| cA01 | 0.90 | 0.10 | 30.0 | 1.35 | | -18% | 77% | 0.03 | 0.02 | 0.12 | 0.00 | 0.00 | 0.67 | 1.04 |
| cA02 | 0.90 | 0.10 | 32.5 | 1.30 | | -26% | 69% | 0.05 | 0.02 | 0.13 | 0.01 | 0.00 | 0.70 | 1.07 |
| cA03 | 0.90 | 0.10 | 32.5 | 1.50 | | -35% | 61% | 0.10 | 0.03 | 0.18 | 0.02 | 0.00 | 0.57 | 1.24 |
| cA04 | 0.90 | 0.10 | 35.0 | 1.30 | | -36% | 60% | 0.16 | 0.05 | 0.23 | 0.06 | 0.00 | 0.68 | 1.14 |
| cA05 | 0.90 | 0.10 | 35.0 | 1.35 | 30 / 70 / 0 | -36% | 61% | 0.20 | 0.06 | 0.26 | 0.07 | 0.00 | 0.64 | 1.19 |
| cA06 | 0.90 | 0.10 | 35.0 | 2.00 | | -31% | 68% | 0.02 | 0.01 | 0.09 | 0.01 | 0.00 | 0.33 | 1.76 |
| cA07 | 0.90 | 0.10 | 45.0 | 1.00 | | -49% | 46% | 0.53 | 0.14 | 0.81 | 0.00 | 0.00 | 0.95 | 1.08 |
| cA08 | 0.90 | 0.10 | 48.0 | 1.00 | | -66% | 31% | 1.50 | 0.28 | 1.27 | 0.00 | 0.00 | 0.97 | 1.14 |
| cA09 | 0.90 | 0.10 | 50.0 | 1.00 | | -66% | 32% | 0.68 | 0.15 | 0.80 | 0.02 | 0.00 | 0.94 | 1.18 |
| cA10 | 0.90 | 0.10 | 53.0 | 1.00 | | -75% | 23% | 0.89 | 0.21 | 0.96 | 0.15 | 0.00 | 0.96 | 1.23 |
| cB01 | 0.90 | 0.15 | 32.5 | 1.15 | | -41% | 53% | 0.10 | 0.03 | 0.23 | 0.00 | 0.00 | 1.06 | 1.49 |
| cB02 | 0.90 | 0.15 | 35.0 | 1.15 | 30 / 70 / 0 | -35% | 58% | 0.23 | 0.06 | 0.37 | 0.01 | 0.00 | 1.10 | 1.59 |
| cB03 | 0.90 | 0.15 | 35.0 | 1.20 | | -33% | 60% | 0.53 | 0.15 | 0.86 | 0.05 | 0.00 | 1.06 | 1.66 |
| cB04 | 0.90 | 0.15 | 40.0 | 1.10 | | -41% | 53% | 1.20 | 0.27 | 1.41 | 0.04 | 0.00 | 1.16 | 1.71 |
| cC01 | 0.90 | 0.20 | 30.0 | 1.30 | | -34% | 57% | 0.52 | 0.16 | 1.00 | 0.06 | 0.00 | 1.20 | 2.18 |
| cC02 | 0.90 | 0.20 | 32.5 | 1.20 | | -30% | 61% | 0.90 | 0.27 | 1.63 | 0.03 | 0.00 | 1.40 | 2.16 |
| cC03 | 0.90 | 0.20 | 32.5 | 1.25 | | -37% | 54% | 1.01 | 0.27 | 1.51 | 0.06 | 0.00 | 1.27 | 2.25 |
| cC04 | 0.90 | 0.20 | 32.5 | 1.30 | 30 / 70 / 0 | -32% | 58% | 1.12 | 0.27 | 1.39 | 0.14 | 0.00 | 1.30 | 2.34 |
| cC05 | 0.90 | 0.20 | 35.0 | 1.15 | | -36% | 54% | 1.32 | 0.35 | 1.98 | 0.03 | 0.00 | 1.46 | 2.21 |
| cC06 | 0.90 | 0.20 | 35.0 | 1.20 | | -35% | 56% | 1.24 | 0.29 | 1.60 | 0.01 | 0.00 | 1.28 | 2.31 |
| cC07 | 0.90 | 0.20 | 45.0 | 1.00 | | -54% | 39% | 1.30 | 0.31 | 1.61 | 0.06 | 0.00 | 1.74 | 2.37 |
| cC08 | 0.90 | 0.20 | 50.0 | 1.00 | | -76% | 20% | 3.16 | 0.65 | 2.84 | 0.37 | 0.00 | 2.02 | 2.57 |
| fA01 | 0.90 | 0.20 | 30.0 | 1.30 | | -28% | 64% | 0.95 | 0.29 | 1.50 | 0.37 | 0.00 | 1.40 | 2.16 |
| fA02 | 0.90 | 0.20 | 35.0 | 1.20 | | -28% | 63% | 1.18 | 0.26 | 1.20 | 0.13 | 0.00 | 1.55 | 2.28 |
| fA03 | 0.90 | 0.20 | 32.5 | 1.25 | 50 / 50 / 0 | -31% | 62% | 1.16 | 0.29 | 1.41 | 0.25 | 0.00 | 1.48 | 2.23 |
| fA04 | 0.90 | 0.20 | 35.0 | 1.25 | | -24% | 67% | 1.33 | 0.29 | 1.22 | 0.26 | 0.00 | 1.57 | 2.38 |
| fA05 | 0.90 | 0.20 | 40.0 | 1.10 | | -33% | 59% | 1.17 | 0.29 | 1.56 | 0.09 | 0.00 | 1.68 | 2.34 |
| fB01 | 0.90 | 0.10 | 30.0 | 1.35 | | -23% | 74% | 0.03 | 0.02 | 0.08 | 0.05 | 0.00 | 0.74 | 1.01 |
| fB02 | 0.90 | 0.10 | 35.0 | 1.30 | | -28% | 69% | 0.38 | 0.11 | 0.37 | 0.30 | 0.00 | 0.78 | 1.12 |
| fB03 | 0.90 | 0.10 | 40.0 | 1.10 | | -25% | 73% | 0.31 | 0.10 | 0.46 | 0.16 | 0.00 | 0.85 | 1.06 |
| fB04 | 0.90 | 0.10 | 45.0 | 1.00 | 70 / 30 / 0 | -19% | 78% | 0.39 | 0.11 | 0.59 | 0.12 | 0.00 | 0.89 | 1.06 |
| fB05 | 0.90 | 0.10 | 48.0 | 1.00 | | -33% | 64% | 0.72 | 0.18 | 0.51 | 0.54 | 0.00 | 0.97 | 1.12 |
| fB06 | 0.90 | 0.20 | 30.0 | 1.30 | | -19% | 75% | 1.48 | 0.37 | 1.40 | 0.68 | 0.00 | 1.47 | 2.13 |
| fB07 | 0.90 | 0.20 | 30.0 | 1.35 | | -18% | 76% | 1.63 | 0.38 | 1.38 | 0.71 | 0.00 | 1.48 | 2.21 |

Table 1.: Simulation results. Note that $L_D$ and $M_{moon}$ estimated according to (Kokubo 2000) considers all disk material, including any present iron and water ice.

| run | $M_{tar}$ | $M_{imp}$ | θ | $v_{imp}$ | impactor | $\delta f_T$ | $f_T$ | $M_{moon}$ | $L_D$ | $M_D$ $[M_L]$ | | | $L_{bound}$ | $L_{imp}$ |
|---|---|---|---|---|---|---|---|---|---|---|---|---|---|---|
| | $[M_E]$ | $[M_E]$ | [°] | $[v_{esc}]$ | [wt% Fe/$SiO_2$/$H_2O$] | | | $[M_L]$ | $[L_{E-M}]$ | $SiO_2$ | Fe | $H_2O$ | $[L_{E-M}]$ | $[L_{E-M}]$ |
| iA01 | 0.90 | 0.20 | 15.0 | 1.50 | | -6% | 84% | -0.13 | 0.02 | 0.21 | 0.00 | 0.08 | 0.75 | 1.35 |
| iA02 | 0.90 | 0.20 | 25.0 | 1.30 | | -26% | 66% | 0.17 | 0.09 | 0.47 | 0.00 | 0.19 | 1.12 | 1.92 |
| iA03 | 0.90 | 0.20 | 25.0 | 1.35 | | -23% | 69% | 0.15 | 0.08 | 0.41 | 0.00 | 0.18 | 1.08 | 1.99 |
| iA04 | 0.90 | 0.20 | 25.0 | 1.50 | | -16% | 76% | 0.07 | 0.07 | 0.47 | 0.00 | 0.16 | 0.92 | 2.21 |
| iA05 | 0.90 | 0.20 | 25.0 | 1.75 | | -24% | 71% | 0.17 | 0.07 | 0.38 | 0.01 | 0.11 | 0.60 | 2.58 |
| iA06 | 0.90 | 0.20 | 30.0 | 1.00 | | -20% | 71% | 0.20 | 0.10 | 0.20 | 0.00 | 0.52 | 1.28 | 1.74 |
| iA07 | 0.90 | 0.20 | 30.0 | 1.15 | | -52% | 43% | 0.76 | 0.21 | 0.62 | 0.00 | 0.65 | 1.36 | 2.01 |
| iA08 | 0.90 | 0.20 | 30.0 | 1.20 | | -62% | 34% | 0.91 | 0.23 | 0.75 | 0.00 | 0.59 | 1.35 | 2.09 |
| iA09 | 0.90 | 0.20 | 30.0 | 1.25 | | -20% | 72% | 0.36 | 0.11 | 0.33 | 0.00 | 0.35 | 1.15 | 2.18 |
| iA10 | 0.90 | 0.20 | 30.0 | 1.30 | | -10% | 81% | 0.60 | 0.17 | 0.73 | 0.00 | 0.33 | 1.12 | 2.27 |
| iA11 | 0.90 | 0.20 | 30.0 | 1.32 | | -14% | 78% | 0.26 | 0.09 | 0.26 | 0.00 | 0.37 | 1.00 | 2.30 |
| iA12 | 0.90 | 0.20 | 30.0 | 1.35 | | -10% | 82% | 0.32 | 0.11 | 0.45 | 0.00 | 0.30 | 0.96 | 2.36 |
| iA13 | 0.90 | 0.20 | 32.5 | 1.25 | | -15% | 77% | 0.71 | 0.19 | 0.56 | 0.00 | 0.56 | 1.01 | 2.34 |
| iA14 | 0.90 | 0.20 | 32.5 | 1.30 | 15 / 35 / 50 | -23% | 70% | 1.08 | 0.23 | 0.79 | 0.00 | 0.36 | 1.09 | 2.44 |
| iA15 | 0.90 | 0.20 | 32.5 | 1.35 | | -54% | 42% | 2.19 | 0.45 | 1.50 | 0.26 | 0.48 | 1.06 | 2.53 |
| iA16 | 0.90 | 0.20 | 35.0 | 1.00 | | -60% | 36% | 1.35 | 0.32 | 0.70 | 0.00 | 1.03 | 1.51 | 2.00 |
| iA17 | 0.90 | 0.20 | 35.0 | 1.10 | | -30% | 63% | 1.61 | 0.37 | 0.92 | 0.00 | 1.07 | 1.42 | 2.20 |
| iA18 | 0.90 | 0.20 | 35.0 | 1.15 | | -60% | 36% | 3.03 | 0.59 | 1.50 | 0.00 | 1.23 | 1.57 | 2.30 |
| iA19 | 0.90 | 0.20 | 35.0 | 1.20 | | -60% | 36% | 2.89 | 0.52 | 1.23 | 0.00 | 0.95 | 1.55 | 2.40 |
| iA20 | 0.90 | 0.20 | 35.0 | 1.25 | | -56% | 40% | 2.26 | 0.50 | 1.73 | 0.16 | 0.74 | 1.23 | 2.50 |
| iA21 | 0.90 | 0.20 | 35.0 | 1.30 | | 3% | 98% | -0.01 | 0.00 | 0.02 | 0.00 | 0.02 | 0.48 | 2.60 |
| iA22 | 0.90 | 0.20 | 40.0 | 1.15 | | -70% | 27% | 8.14 | 1.39 | 2.80 | 0.80 | 1.97 | 2.03 | 2.58 |
| iA23 | 0.90 | 0.20 | 45.0 | 1.00 | | -67% | 30% | 2.04 | 0.49 | 1.28 | 0.01 | 1.39 | 1.71 | 2.47 |
| iA24 | 0.90 | 0.20 | 45.0 | 1.15 | | -61% | 37% | 0.09 | 0.04 | 0.09 | 0.00 | 0.19 | 1.14 | 2.84 |
| iA25 | 0.90 | 0.20 | 45.0 | 1.20 | | -1% | 97% | 0.01 | 0.01 | 0.03 | 0.00 | 0.05 | 0.35 | 2.96 |
| iA26 | 0.90 | 0.20 | 45.0 | 1.25 | | -10% | 89% | 0.01 | 0.01 | 0.03 | 0.00 | 0.04 | 0.35 | 3.08 |
| iA27 | 0.90 | 0.20 | 60.0 | 1.00 | | -73% | 24% | 0.96 | 0.28 | 0.75 | 0.01 | 0.99 | 1.41 | 3.02 |
| iA28 | 0.90 | 0.20 | 60.0 | 1.15 | | 0% | 100% | 0.00 | 0.00 | 0.01 | 0.00 | 0.02 | 0.16 | 3.48 |

Table 1. (continued)